\documentclass[a4paper,12pt]{article}
\usepackage{a4wide}
\usepackage{amsmath}
\usepackage{amsfonts}
\usepackage{bbm}
\newcommand{\be}{\begin{eqnarray}}
\newcommand{\beq}{\begin{eqnarray}}
\newcommand{\ee}{\end{eqnarray}}
\newcommand{\tx}{\textrm}
\newcommand{\ov}{\overline}
\newcommand{\p}{\partial}

\newcommand{\cx}{\mathcal}
\newcommand{\Ga}{\Gamma}
\newcommand{\ib}{\ov{i}}

\newcommand{\bb}[1][\overline]{1}

\newcommand{\Fgh}{\mathcal{F}^{(g,h)}}
\newcommand{\ue}{\text{e}}
\newcommand{\h}{\frac{1}{2}}
\newcommand{\deltanew}{\mathcal{E}}

\begin{document}

\setlength{\parindent}{0cm}
\setlength{\baselineskip}{1.5em}

\title{\bf Polynomial Structure of the (Open) Topological String Partition Function}
\author{Murad Alim$^{a}$\footnote{\tt{alim@theorie.physik.uni-muenchen.de}} and Jean Dominique L\"ange$^{a}$\footnote{\tt{jdl@theorie.physik.uni-muenchen.de}}\\[20pt]
\it $^{a}$Arnold Sommerfeld Center for Theoretical Physics\\
\it Ludwig-Maximilians-University \\
\it Department of Physics \\
\it Theresienstr. 37, D-80333 M\"unchen, Germany}
\date{}
\maketitle

\vspace{-250pt}
\hfill{LMU-ASC 57/07}
\vspace{+250pt}

\begin{center}
{\bf Abstract}
\end{center}
In this paper we show that the polynomial structure of the topological string partition function found by Yamaguchi and Yau for the quintic holds for  an arbitrary Calabi-Yau manifold with any number of moduli. Furthermore, we generalize these results to the open topological string partition function as discussed recently by Walcher and reproduce his results for the real quintic.

\clearpage

\section{Introduction and Summary}
The holomorphic anomaly equation of the topological string \cite{BCOV1,BCOV2} relates the anti-holomorphic derivative of the genus $g$ topological string partition function $\mathcal{F}^{(g)}$ with covariant derivatives of the partition functions of lower genus. This enables one to recursively determine the partition function at each genus up to a holomorphic ambiguity which has to be fixed by further information. A complete understanding of the holomorphic anomaly equation and its recursive procedure to determine the partition functions at every genus might lead to new insights in the understanding of the structure of the full topological string partition function $Z=\exp(\sum \lambda^{2g-2}\mathcal{F}^{(g)})$. For example in \cite{Witten}, Witten interpreted $Z$ as a wave function for the quantization of the space $\text{H}^3(X,\mathbbm{R})$ of a Calabi-Yau $X$ and the holomorphic anomaly equation as the background independence of this wave function.
In \cite{Yau}, Yamaguchi and Yau discovered that the non-holomorphic part of the topological string partition function for the quintic can be written as a polynomial in a finite number of generators. This improves the method using Feynman rules proposed in \cite{BCOV2}. This polynomial structure was used in \cite{HKQ} to solve the quintic up to genus 51 and was applied to other Calabi-Yau manifolds with one modulus in \cite{HKQ,Hosono:2007vf}.

The first aim of this paper is to generalize the polynomial structure of the topological string partition function discovered in \cite{Yau} to an arbitrary Calabi-Yau manifold with any number of moduli.\footnote{This problem has independently solved in \cite{Mayr}.}  A related method for integrating the holomorphic anomaly equation using modular functions was presented in \cite{Aganagic,Grimm}.

Recently, an extension of the holomorphic anomaly equation which includes the open topological string was proposed by Walcher \cite{Walcher1}. Its solution in terms of Feynman rules was proven soon after in \cite{Cook}. The second task of this paper is to extend Yamaguchi and Yau's polynomial construction to the open topological string. We recently learned at the Simons Workshop in Mathematics and Physics 2007 that a similar generalization for the open topological string on the quintic will appear in \cite{Konishi}.

The organisation of the paper is as follows. In the next section we briefly review the extended holomorphic anomaly equation and the initial correlation functions at low genus and number of holes which will be the starting point of the recursive procedure. Next we introduce the polynomial generators of the non-holomorphic part of the partition functions and show that holomorphic derivatives thereof can again be expressed in terms of these generators. As the initial correlation functions are expressions in these generators we will have thus shown that at every genus the partition functions will be again expressions in the generators. Afterwards we assign some grading to the generators and show that $\mathcal{F}^{(g,h)}_{i_1\dots i_n}$, the partition function at genus $g$, with $h$ holes and $n$ insertions, will be a polynomial of degree $3g-3+3h/2+n$ in the generators. Finally, we determine the polynomial recursion relations and argue that, by a change of generators, the number of generators can be reduced by one.
In order to solve the holomorphic anomaly equation it now suffices to make the most general Ansatz of the right degree in the generators for the partition function and use the recursion relation to match the coefficients.
This procedure allows to determine the partition function up to some holomorphic ambiguity in every step. In the third section we apply our method to the real quintic and give the polynomial expressions for the partition functions and reproduce some recent results.

Some subtleties of our approach still require further investigations, most of these are related to parametrizing the holomorphic ambiguities. There is a freedom in determining the holomorphic part of the generators which changes the complexity of the holomorphic ambiguity at every step. For the closed string part of the quintic we fixed the holomorphic part of the generators as in \cite{Hosono}, the ambiguities in the partition functions are then polynomials in the inverse discriminant. The Ansatz for these polynomials can be deduced in order to reproduce the right behaviour of the partition function at special points in the moduli space. It would be interesting to further understand the structure of the holomorphic part of the partition function and find out whether there is some systematic way to completely determine the topological string partition function.

After we finished this paper a generalization of the holomorphic anomaly equations for the open topological string appeared in \cite{Bonelli}.

\section{Polynomial Structure of Topological String Partition Functions}

\subsection{Holomorphic Anomaly}
In this paper we consider the open topological string with branes as in \cite{Walcher1}.
The B-model on a Calabi--Yau manifold $X$ depends on the space 
$\mathcal M$ of complex structures parametrized by coordinates
$z^i$, $i=1,...,h^{1,2}(X)$. More precisely, the topological string partition function $\Fgh$
at genus $g$ with $h$ boundaries
is a section of a line bundle $\mathcal{L}^{2-2g-h}$ over $\mathcal M$ \cite{Walcher1}.
The line bundle $\mathcal L$ may be
identified with the bundle of holomorphic $(3,0)$-forms $\Omega$ on $X$
with first Chern class $G_{i\bar{j}}=\partial_i \bar{\partial}_{\bar{j}}
K$ . Here $K$ is the K\"ahler potential
and $G_{i\bar{j}}$ the K\"ahler metric. Under K\"ahler transformations 
$K \rightarrow K(z^i,\bar{z}^{\bar{j}})-\ln \phi(z^i)-\ln
\bar{\phi}(\bar{z}^{\bar{j}})$,
$\Omega\rightarrow \phi\Omega$ and more generally 
a section $f$ of $\mathcal{L}^n \otimes \bar{\mathcal{L}}^{\bar{n}}$
transforms as
$f \rightarrow \phi^n \bar{\phi}^{\bar{n}}f$. 

The fundamental objects of the topological string are the holomorphic
three point couplings
at genus zero $C_{ijk}$ which can be
integrated to the genus zero partition function $F_0$
\begin{eqnarray}
C_{ijk}=D_i D_j D_k F_0, \qquad \bar{\partial}_{\bar{i}} C_{ijk}=0
\label{feyn1}
\end{eqnarray}
and the disk amplitudes with two bulk insertions
$\Delta_{ij}$ which are symmetric in the
two indices but not holomorphic
\begin{eqnarray}
\bar{\partial}_{\bar{i}} \Delta_{ij} = - C_{ijk} \Delta^k_{\bar{i}},
\qquad \Delta_{\bar{i}}^k = \Delta_{\bar{i} \bar{j}} \ue^K G^{k
\bar{j}}. \label{feyn2}
\end{eqnarray}
Here $\Delta_{\bar{i}\bar{j}}$ denotes the complex conjugate of
$\Delta_{ij}$ and
$ D_i=\partial_i+\dots =\frac{\partial}{\partial z_i}+ \dots $ 
denotes the covariant derivative on the bundle $\cx L^m\otimes 
\tx{Sym}^n T^*$ where $m$ and $n$ follow from the context. $T^*$ is
the cotangent bundle of $\cx M$ with the standard connection coefficients
$\Ga^i_{jk}=G^{i\ib}\p_jG_{k\ib}$. The connection on the bundle $\cx L$ is
given by the first derivatives of the K\"ahler potential $K_i=\p_iK$.\footnote{See section two of
\cite{BCOV2} for further background material.}

The correlation function at genus $g$ with $h$ boundaries and $n$
insertions $\mathcal{F}^{(g,h)}_{i_1\cdots i_n}$ is only non-vanishing for
$(2g-2+h+n)>0$. They are related by taking covariant derivatives as this
represents insertions of chiral operators in the bulk, 
e.g. 
$D_i \mathcal{F}^{(g,h)}_{i_1\cdots i_n}=\mathcal{F}^{(g,h)}_{ii_1\cdots i_n}$.

Furthermore, in \cite{Walcher1} it is shown that the genus $g$ partition
function with $h$ holes is recursively related to lower genus partition
functions and to partition functions with less boundaries.
This is expressed  for $(2g-2+h)>0$ by an extension of the holomorphic anomaly equations of BCOV \cite{BCOV2}
\begin{eqnarray}
\bar{\partial}_{\bar{i}} \Fgh = \h \bar{C}_{\bar{i}}^{jk}
\sum_{\genfrac{}{}{0pt}{}{g_1+g_2=g}{h_1+h_2=h}}
D_j\mathcal{F}^{(g_1,h_1)} D_k\mathcal{F}^{(g_2,h_2)} + \h
\bar{C}^{jk}_{\bar{i}} D_jD_k\mathcal{F}^{(g-1,h)} - \Delta^j_{\bar{i}}
D_j \mathcal{F}^{(g,h-1)} \label{feyn3}
\end{eqnarray}
where
\begin{eqnarray}
\bar{C}_{\bar{k}}^{ij}= \bar{C}_{\bar{i} \bar{j}\bar{k}} G^{i
\bar{i}}G^{j \bar{j}}\, \ue^{2K}, \qquad \bar{C}_{\bar{i}\bar{j}\bar{k}}=
\overline{C_{ijk}}. \label{feyn4}
\end{eqnarray}
These equations, supplemented by
\begin{eqnarray}
\bar{\partial}_{\bar{i}} \mathcal{F}^{(1,0)}_j &=& \frac{1}{2} C_{jkl}
C^{kl}_{\bar{i}}+ (1-\frac{\chi}{24})
G_{j \bar{i}}, \label{feyn5} \\
\bar{\partial}_{\bar{i}} \mathcal{F}^{(0,2)}_j &=& - \Delta_{jk}
\Delta^k_{\bar{i}} + \frac{N}{2} G_{j\bar{i}} \label{feyn6}
\end{eqnarray}
and special geometry, determine all correlation functions up
to holomorhpic ambiguities. In (\ref{feyn5}), $\chi$ is the Euler character of the manifold and in (\ref{feyn6}) $N$ is the rank
of a bundle over $\mathcal{M}$ in which the charge zero ground states of
the open string live.
Similar to the closed topological string \cite{BCOV2}, a solution of the
recursion equations is given in terms of Feynman rules. These Feynman
rules have been proven for the open topological string in \cite{Cook}.

The propagators for these Feynman rules contain the ones already present for
the closed topological string
$S$, $S^i$, $S^{ij}$ and new propagators $\Delta$, $\Delta^i$. Note that
these are not the same as the $\Delta$, with or without indices, that appear
in \cite{BCOV2} which there denote the inverses of the $S$ propagators. $S$, $S^i$ and $S^{ij}$
are related to the three point couplings $C_{ijk}$ as
\begin{eqnarray}
\partial_{\bar{i}} S^{ij}= \bar{C}_{\bar{i}}^{ij}, \qquad
\partial_{\bar{i}} S^j = G_{i\bar{i}} S^{ij}, \qquad
\partial_{\bar{i}} S = G_{i \bar{i}} S^i.
\label{feyn7}
\end{eqnarray}
By definition, the propagators $S$, $S^i$ and $S^{ij}$ are sections of
the bundles
$\mathcal{L}^{-2}\otimes \text{Sym}^m T$ with $m=0,1,2$.
$\Delta$ and $\Delta^i$ are related to the disk amplitudes with two
insertions by
\begin{eqnarray}
\bar{\partial}_{\bar{i}} \Delta^j = \Delta^j_{\bar{i}}, \qquad
\bar{\partial}_{\bar{i}} \Delta = G_{i\bar{i}} \Delta^i.
\label{feyn8}
\end{eqnarray}
They are sections of $\mathcal{L}^{-1}\otimes \text{Sym}^m T$ with
$m=0,1$. The vertices of the Feynman rules are
given by the correlation functions $\mathcal{F}^{(g,h)}_{i_1\cdots
i_n}$.

Note that the anomaly equation (\ref{feyn3}), as well as the definitions
(\ref{feyn7}) and (\ref{feyn8}), leave the freedom of adding holomorphic
functions under the
$\ov{\partial}$ derivatives as integration constants. This freedom is referred
to as holomorphic ambiguities.

\subsection{Initial Correlation Functions}
To be able to apply a recursive procedure for solving the holomorphic anomaly equation, we first need to have some initial data to start with. In this case the initial data consists of the first non-vanishing correlation functions. The first non-vanishing correlation functions
at genus zero without any boundaries are the holomorphic three point couplings
$\mathcal{F}^{(0,0)}_{ijk}\equiv C_{ijk}$. At genus zero with one
boundary, the first non-vanishing correlation functions are the disk
amplitudes with two insertions. The holomorhpic anomaly equation
(\ref{feyn2}) is solved with (\ref{feyn8}) by
\begin{eqnarray}
\mathcal{F}^{(0,1)}_{ij} \equiv \Delta_{ij} = - C_{ijk} \Delta^k +
g_{ij}
\label{sol1}
\end{eqnarray}
with some holomorphic functions $g_{ij}$. Finally we solve
(\ref{feyn5}) and (\ref{feyn6}). (\ref{feyn5}) can be integrated wih
(\ref{feyn7}) to
\begin{eqnarray}
\mathcal{F}^{(1,0)}_i = \h C_{ijk} S^{jk} +(1-\frac{\chi}{24}) K_i
+ f_i^{(1,0)} \label{sol2}
\end{eqnarray}
with ambiguity $f_i^{(1,0)}$. For the annulus we find
\begin{eqnarray}
\bar{\partial}_{\bar{i}} \mathcal{F}^{(0,2)}_j &=& C_{jkl} \Delta^l
\bar{\partial}_{\bar{i}} \Delta^k + \bar{\partial}_{\bar{i}} (- g_{jk}
\Delta^k + \frac{N}{2} K_j) \nonumber \\
&=& \bar{\partial}_{\bar{i}} ( \h C_{jkl} \Delta^k \Delta^l - g_{jk}
\Delta^k + \frac{N}{2} K_j)
\label{sol3}
\end{eqnarray}
and therefore
\begin{eqnarray}
\mathcal{F}^{(0,2)}_i = \h C_{ijk} \Delta^j \Delta^k - g_{ij} \Delta^j +
\frac{N}{2} K_i + f^{(0,2)}_i
\label{sol4}
\end{eqnarray}
where $f^{(0,2)}_i$ are holomorhpic. As can be seen from
these expressions, the non-holomorphicity of the correlation functions
only comes from the propagators together with $K_i$. Indeed, we will
now show that this holds for all partition functions $\Fgh$.

\subsection{Non-holomorphic Generators}
From the holomorphic anomaly equation and its Feynman rule solution it
is clear that at every genus $g$ with $h$ boundaries the building blocks
of the partition function $\Fgh$ are the propagators $S^{ij}$, $S^i$,
$S$, $\Delta$, $\Delta^i$ and vertices $\mathcal{F}^{(g',h')}_{i_1\cdots i_n}$ with $g'<g$ or $h'< h$. Here it will be shown that all the
non-holomorphic content of the partition functions $\Fgh$ can be expressed in terms of
a finite number of generators.
The generators we consider are the propagators
$S^{ij}, S^i$, $S$, $\Delta^i$, $\Delta$ as well as $K_i$, the partial
derivative of the
K\"ahler potential. This construction is a generalization of Yamaguchi and Yau's
polynomial construction for the quintic \cite{Yau} where multi
derivatives of the connections were used as generators.
The propagators of the closed topological string as building
blocks were also used recently by
Grimm, Klemm, Marino and Weiss \cite{Grimm} for a
direct integration of the topological string using modular properties of the
big moduli space, where all propagators can be treated on equal footing.

In the following we prove that if the anti-holomorphic part of $\Fgh$ is
expressed in terms of the generators $S^{ij}$, $S^i$, $S$, $\Delta^i$,
$\Delta$ and $K_i$, then all covariant derivatives thereof are also
expressed in terms of these generators. As the correlation functions for
small genus and small number of boundaries are expressed in terms of the
generators, it follows by induction, that all $\Fgh$ are expressed
in terms of the generators.

The covariant derivatives contain the Christoffel connection and
the connection $K_i$ of $\mathcal{L}$. By integrating the
special geometry relation
\begin{eqnarray}
\bar{\partial}_{\bar{i}} \Gamma^l_{ij}= \delta_i^l
G_{j\bar{i}} + \delta_j^l G_{i\bar{i}} - C_{ijk} C^{kl}_{\bar{i}}
\label{finite5}
\end{eqnarray}
to
\begin{eqnarray}
\Gamma^l_{ij} = \delta_i^l K_j + \delta^l_j K_i - C_{ijk} S^{kl} + s^l_{ij},
\label{finite6}
\end{eqnarray}
where $s^l_{ij}$ denote holomorphic functions that are not fixed by the
special geometry relation, we can express the Christoffel connection in
terms of our generators. What remains is to show that the covariant
derivatives of all generators are again expressed in terms of the generators.
To obtain expressions for the covariant
derivatives of the generators we first take the anti-holomorphic
derivative of the expression, then use (\ref{finite5}) and write the
result as a total anti-holomorphic derivative again, for
example
\begin{eqnarray}
\partial_{\bar{i}} (D_i S^{jk})= \partial_{\bar{i}}(\delta_i^j S^k +
\delta_i^k S^j - C_{imn} S^{mj} S^{nk}).
\label{finite8}
\end{eqnarray}
This equation determines $D_i S^{jk}$ up to a holomorphic term.
In this manner we obtain the following relations
\begin{eqnarray}
D_i S^{jk} &=& \delta_i^j  S^k + \delta_i^k S^j - C_{imn}S^{mj} S^{nk} +
h_i^{jk} \label{finite9}, \\
D_i S^j &=& 2 \delta_i^j S - C_{imn} S^m S^{nj} + h_i^{jk} K_k  + h_i^j,
\label{finite10} \\
D_i S &=& -\frac{1}{2} C_{imn} S^m S^n + \frac{1}{2} h^{mn}_i K_m K_n +
h_i^j K_j + h_i, \label{finite11} \\
D_i K_j &=& -K_i K_j - C_{ijk} S^k + C_{ijk} S^{kl} K_l + h_{ij},
\label{finite12} \\
D_i \Delta^j &=& \delta_i^j \Delta - g_{ik} S^{kj} + g_i^j, \label{finite13} \\
D_i \Delta &=& - g_{ij} S^j + g_i^j K_j + g_i, \label{finite14}
\end{eqnarray}
where $h_i^{jk}, h^j_i$, $h_i$, $h_{ij}$, $g_i^j$ and $g_i$ denote holomorphic functions
(ambiguities). This completes our
proof that all non-holomorphic parts of $\Fgh$ can be expressed in terms
of the generators. Next, we will determine recursion relations, asign
some grading to the generators and show that $\Fgh_{i_1\cdots i_n}$ is
a polynomial of degree $3g-3+3h/2 + n$.

\subsection{Polynomial Recursion Relation}
Let us now determine some recursion relations from the holomorhpic
anomaly equation.
Computing the $\bar{\partial}_{\bar{i}}$ derivative of
$\Fgh$ expressed in terms of  $S^{ij}$, $S^i$, $S$, $\Delta^i$, $\Delta$,
$K_i$, and using (\ref{feyn3}) one obtains
\begin{eqnarray}
&& \bar{C}_{\bar{i}}^{jk} \frac{\partial \Fgh}{\partial S^{jk}} +
\Delta_{\bar{i}}^j \frac{\partial \Fgh}{\partial
\Delta^j} + G_{i\bar{i}} \left(
\frac{\partial \Fgh}{\partial K_i} + S^i \frac{\partial \Fgh}{\partial
S} + S^{ij} \frac{\partial \Fgh}{\partial S^j} + \Delta^i \frac{\partial
\Fgh}{\partial \Delta}
 \right) \nonumber \\
&=& \h \bar{C}_{\bar{i}}^{jk}
\sum_{\genfrac{}{}{0pt}{}{g_1+g_2=g}{h_1+h_2=h}}
D_j\mathcal{F}^{(g_1,h_1)} D_k\mathcal{F}^{(g_2,h_2)} + \h
\bar{C}^{jk}_{\bar{i}} D_jD_k\mathcal{F}^{(g-1,h)} - \Delta^j_{\bar{i}}
D_j \mathcal{F}^{(g,h-1)}. \label{rec1}
\end{eqnarray}
Assuming linear independence of $\bar{C}^{jk}_{\bar{i}}$,
$\Delta^j_{\bar{i}}$ and $G_{i\bar{i}}$
the equation splits into three equations
\begin{eqnarray}
\frac{\partial \Fgh}{\partial S^{ij}} &=&  \h
\sum_{\genfrac{}{}{0pt}{}{g_1+g_2=g}{h_1+h_2=h}}
D_i\mathcal{F}^{(g_1,h_1)} D_j\mathcal{F}^{(g_2,h_2)} + \h
D_iD_j\mathcal{F}^{(g-1,h)}, \label{rec2}
\\
\frac{\partial \Fgh}{\partial \Delta^i} &=& - D_i \mathcal{F}^{(g,h-1)},
\label{rec3} \\
0 &=& \frac{\partial \Fgh}{\partial K_i} + S^i \frac{\partial \Fgh}{\partial
S} + S^{ij} \frac{\partial \Fgh}{\partial S^j} + \Delta^i
\frac{\partial \Fgh}{\partial \Delta}. \label{rec4}
\end{eqnarray}
The last equation (\ref{rec4}) can be rephrased as the condition that
$\Fgh$ does not depend explicitly on $K_i$ by making a suitable change of
generators
\begin{eqnarray}
\tilde{S}^{ij} &=& S^{ij}, \label{rec5} \\
\tilde{S}^i &=& S^i - S^{ij} K_j, \label{rec6} \\
\tilde{S} &=& S- S^i K_i + \frac{1}{2} S^{ij} K_i K_j, \label{rec7}\\
\tilde{\Delta}^i &=& \Delta^i, \label{rec8} \\
\tilde{\Delta} &=& \Delta - \Delta^i K_i, \label{rec9} \\
\tilde{K_i} &=& K_i, \label{rec10}
\end{eqnarray}
i.e. $\partial \Fgh / \partial \tilde{K}_i =0$ for $\Fgh$ as a function
of the tilded generators. Let us now asign a grading to the generators
and covariant derivatives, which is naturally
inherited from the $U(1)$ grading given by the background charge for the
$U(1)$ current inside the twisted $\mathcal{N}=2$ superconformal
algebra. The covariant holomorphic derivatives $D_i$ carry charge +1 as
they represent the insertion of a chrial operator of $U(1)$ charge
+1. As $K_i$ is part of the connection, it is natural to asign charge
+1 to $K_i$. From the definitions (\ref{feyn7}) and (\ref{feyn8}) one may asign
the charges $1/2,1,3/2,2,3$ to the generators $\Delta^i$, $S^{ij}$,
$\Delta$, $S^i$, $S$, respectively. The correlation functions
$\Fgh_{i_1\cdots i_n}$ for small
$g$ and $h$ are a polynomial of degree $3g-3+3h/2+n$ in the generators.
By the recursion relations, it immediately follows that this holds for
all $g$ and $h$.

\section{The Real Quintic}
As an example of our polynomial construction of the partition functions
$\Fgh$ we consider the real quintic
\begin{eqnarray*}
X:=\{P(x)=0\} \subset \mathbb{P}^4
\end{eqnarray*} 
where $P$ is a homogeneous polynomial of degree 5 in 5 variables $x_1, \dots ,x_5$ with real coefficients. The real locus
\begin{eqnarray*}
L=\{x_i=\bar{x}_i\}
\end{eqnarray*}
is a Lagrangian submanifold on which the boundary of the Riemann surface can be mapped.

For the closed topological string the polynomial construction was discovered
by Yamaguchi and Yau in \cite{Yau} and has been used in \cite{HKQ} to calculate
$\mathcal{F}^{(g,0)}$ up to $g=51$.
The open string case was analyzed in \cite{Walcher1,Walcher2} where the real quintic is given as an example for solving the extended holomorphic anomaly equation. We will follow the notation of these two papers.

The mirror quintic has one complex
structure modulus, which will be denoted by $z$. To parametrize the holomorphic ambiguities we introduce as a holomorphic generator the inverse of the disrciminant
\begin{eqnarray}
P = \frac{1}{1-5^5z}.
\end{eqnarray}
The Yukawa coupling is given by
\begin{eqnarray}
C_{zzz} = 5 P/z^3. \label{quintic2}
\end{eqnarray}
For computational convenience we use instead of the generators $S^{zz}$, $S^z$, $S$, $\Delta^z$ and $\Delta$
the generators
\begin{eqnarray}
T^{zz}= 5 P  \frac{S^{zz}}{z^2}, \quad T^z=5P \frac{S^z}{z}, \quad T=
5P S, \quad \deltanew^z= P^{1/2} \frac{\Delta^z}{z} \quad \text{and} \quad \deltanew = P^{1/2} \Delta.
\end{eqnarray}
To obtain explicit forms of the generators we start with the integrated special geometry relation (\ref{finite6}) and choose similar to \cite{Hosono}
\begin{eqnarray}
s_{zz}^z=-1/z
\end{eqnarray}
in order to cancel the singular term in the holomorhpic limit of $\Gamma^z_{zz}$. In the language of \cite{BCOV2} this corresponds to a gauge choice of $f=z^{-1/2}$ and $v=1$. This choice of holomorhpic ambiguities fixes the propagators $T^{zz}$, $T^z$ and $T$ as
\begin{eqnarray}
T^{zz} &=& 2 \theta K - z \Gamma^z_{zz} -1, \\
T^z &=& (\theta K)^2 - \theta^2 K -\frac{1}{4}, \\
T &=& \left( \frac{1}{5} P - \frac{9}{20}\right) \left(\theta K
-\frac{1}{2}\right) + \frac{1}{2} \left( \theta T^z -(P-1)T^z \right),
\end{eqnarray}
with $\theta=z \frac{\partial}{\partial z}$. This choice of generators leads to the following ambiguities in the derivative relations of the generators (\ref{finite9})-(\ref{finite12})
\begin{eqnarray}
5 P h_z^{zz}/z &=& - \frac{2}{5} P + \frac{9}{10}, \\
5 P h_z^z &=& \frac{1}{5} P - \frac{9}{20}, \\
5 P z h_z &=& - \frac{101}{1250} P + \frac{2241}{20000}, \\
z^2 h_{zz} &=& -\frac{1}{4}.
\end{eqnarray}
For the open string generators $\deltanew^z$ and $\deltanew$ we make the same choice as in \cite{Walcher1} by setting
\begin{eqnarray}
g_{zz} &=& 0 \qquad \text{and} \\
g_z^z &=& 0,
\end{eqnarray}
which leads to
\begin{eqnarray}
\deltanew^z &=& - \frac{1}{5} P^{-1/2} z^2 \Delta_{zz}, \\
\deltanew &=& -\frac{1}{2} (P-1) \deltanew^z + \theta \deltanew^z- T^{zz} \deltanew^z +(\theta K)\deltanew^z.
\end{eqnarray}
Finally, taking the holomorphic limit of (\ref{finite14}) we obtain the last ambiguity in the derivative relations
\begin{eqnarray}
zg_z &=& - \frac{3}{4} z^{1/2}.
\end{eqnarray}
Next, we fix the ambiguities for the initial correlation functions (\ref{sol1}), (\ref{sol2}) and (\ref{sol4}) as in \cite{Walcher1} and obtain
\begin{eqnarray}
z^2 \mathcal{F}^{(0,1)}_{zz} &=& -5 P^{1/2} \deltanew^z, \\
z \mathcal{F}^{(1,0)}_z &=& \frac{28}{3} \theta K + \frac{1}{2}
T^{zz}+\frac{1}{12}P - \frac{13}{6}, \\
z \mathcal{F}^{(0,2)}_z &=& \frac{5 (\deltanew^z)^2}{2} +
\frac{\theta K}{2} + \frac{3 P}{250}-\frac{3}{250}.
\end{eqnarray}
It is now straightforward to use our method to determine higher $\Fgh$ by writing the most general polynomial of degree $3g-3+3h/2$ in the generators $\tilde{T}^{zz}$, $\tilde{T}^z$, $\tilde{T}$, $\tilde{\deltanew}^z$ and $\tilde{\deltanew}$ and using the polynomial recursion relations. For $\mathcal{F}^{(2,0)}$ and $\mathcal{F}^{(3,0)}$ the gap condition at the conifold point \cite{HKQ}  and the known expressions for the contribution of constant maps is enough to fix the holomorphic ambiguities and we give the explicit expressions in Appendix \ref{apppol}. For $\mathcal{F}^{(1,1)}$ and $\mathcal{F}^{(0,3)}$ the vanishing of the first two instanton numbers fixes the ambiguities and read
\begin{eqnarray}
\mathcal{F}^{(1,1)} &=& \frac{28 \tilde{\deltanew}}{3\sqrt{P}} + \frac{13
\tilde{\deltanew}^z}{6\sqrt{P}} - \frac{\tilde{\deltanew}^z \sqrt{P}}{12} -
\frac{\tilde{\deltanew}^z \tilde{T}^{zz}}{2\sqrt{P}}- \frac{9 \sqrt{z} P}{40} +\frac{211\sqrt{z}}{10}, \\
\mathcal{F}^{(0,3)} &=&\frac{1887 \sqrt{z}}{2500} + \frac{\tilde{\deltanew}}{2\sqrt{P}} + \frac{3
\tilde{\deltanew}^z}{250\sqrt{P}} - \frac{5(\tilde{\deltanew}^z)^3}{6\sqrt{P}} - \frac{3 \tilde{\deltanew}^z \sqrt{P}}{250} -
\frac{3\sqrt{z} P}{625}.
\end{eqnarray} 
In Appendix \ref{apppol} we also give the solution of $\mathcal{F}^{(1,2)}$ and $\mathcal{F}^{(2,1)}$ up to the holomorphic ambiguities. It would be interesting to fix this ambiguities by some further input.
\\
\\
{\bf Acknowledgments}\\
We are indebted to P.~Mayr for suggesting the idea of this paper and for many discussions and continuous support. Furthermore we would like to thank I.~Sachs and J.~Walcher for helpful discussions. Finally, J.D.L. is grateful to the Simons Workshop in Mathematics and Physics 2007 for its stimulating atmosphere. The work of M.A. is supported by the ``Studienf\"orderwerk Klaus Murmann" and by the German Excellence Initiative via the programm ``Origin and Structure of the Universe". J.D.L. is supported by a DFG Fellowship with contract number LA 1979/1-1 and the SPP-1096 of the DFG.

\appendix
\section{The Polynomials} 
\label{apppol}
Using the method described in this work we obtained polynomial expression for the topological string partition functions. In this appendix we give the explicit expressions of some of these polynomials in terms of the transformed generators.
\begin{eqnarray}
\mathcal{F}^{(2,0)}&=&  -\frac{1473}{2000} - \frac{139}{375 P} - \frac{43 P}{9000}
+\frac{P^2}{1200} + \frac{140 \tilde{T}}{9 P} - \frac{5 \tilde{T}^z}{36} + \frac{65
\tilde{T}^z}{18 P}- \frac{29 \tilde{T}^{zz}}{ 450} \nonumber \\
&&+ \frac{253\tilde{T}^{zz}}{900P} + \frac{13 P\tilde{T}^{zz}}{1440} - \frac{5 \tilde{T}^z
\tilde{T}^{zz}}{6 P} + \frac{(\tilde{T}^{zz})^2}{30} - \frac{29 (\tilde{T}^{zz})^2}{120 P} +
\frac{(\tilde{T}^{zz})^3}{24 P} 
\end{eqnarray}
\scriptsize
\begin{eqnarray} 
\mathcal{F}^{(3,0)} &&=-\frac{2507719933}{22680000000} - \frac{1208767}{30000000 P^2} - 
    \frac{10405909}{90000000 P} - \frac{1936909 P}{2835000000} +
    \frac{4661 P^2}{5040000} - \frac{29 P^3}{90000} \nonumber\\
    &&+\frac{P^4}{25200}  
    + \frac{2021 \tilde{T}}{67500} + \frac{13066\tilde{T}}{5625 P^2} + \frac{23077
    \tilde{T}}{5000 P} - \frac{47 P\tilde{T}}{9000} -\frac{1316 \tilde{T}^2}{27 P^2} -
    \frac{12319 \tilde{T}^z}{360000} + \frac{14437 \tilde{T}^z}{45000 P^2} \nonumber\\
    &&+ \frac{26201 \tilde{T}^z}{27000 P} + \frac{1067 P \tilde{T}^{z}}{90000} -
    \frac{P^2 \tilde{T}^{z}}{480}
    - \frac{611 \tilde{T} \tilde{T}^{z}}{27 P^2} + \frac{47 \tilde{T} \tilde{T}^{z}}{54 P} + \frac{1603
    (\tilde{T}^{z})^2}{21600} - \frac{105539 (\tilde{T}^{z})^2}{27000 P^2} \nonumber\\
    &&
    - \frac{2621 (\tilde{T}^{z})^2}{27000 P} - \frac{209(\tilde{T}^{z})^3}{81 P^2} -
    \frac{7573 \tilde{T}^{zz}}{720000} - \frac{10231 \tilde{T}^{zz}}{360000P^2}
    + \frac{118493 \tilde{T}^{zz}}{2160000 P} + \frac{48631 P \tilde{T}^{zz}}{4320000}
    \nonumber\\
    &&- \frac{4453 P^2 \tilde{T}^{zz}}{1080000} + \frac{19 P^3 \tilde{T}^{zz}}{36000} -
    \frac{611 \tilde{T} \tilde{T}^{zz}}{10800} - \frac{11891 \tilde{T} \tilde{T}^{zz}}{6750 P^2} +
    \frac{1363 \tilde{T} \tilde{T}^{zz}}{3375P} + \frac{2547 \tilde{T}^{z} \tilde{T}^{zz}}{20000}
    \nonumber\\
    &&- \frac{187013\tilde{T}^{z}\ \tilde{T}^{zz}}{135000 P^2} -\frac{30983 \tilde{T}^{z}
    \tilde{T}^{zz}}{135000 P} - \frac{1613 P \tilde{T}^{z} \tilde{T}^{zz}}{72000} + \frac{47 \tilde{T}
    \tilde{T}^{z}\ \tilde{T}^{zz}}{9 P^2} - \frac{3997 (\tilde{T}^{z})^2 \tilde{T}^{zz}}{2700 P^2}
    \nonumber\\
    &&+ \frac{2719 \tilde{T}^{z}\tilde{T}^{zz}}{5400 P} +
    \frac{61019(\tilde{T}^{zz})^2}{1080000} - \frac{385429(\tilde{T}^{zz})^2}{2160000
    P^2} - \frac{15577 (\tilde{T}^{zz})^2}{360000 P} - \frac{48557 P
    (\tilde{T}^{zz})^2}{2160000}
    \nonumber\\ 
    &&+ \frac{1307 P^2 (\tilde{T}^{zz})^2}{432000} + \frac{1363 \tilde{T}
    (\tilde{T}^{zz})^2}{900 P^2} - \frac{47 \tilde{T} (\tilde{T}^{zz})^2}{225 P} - \frac{251
    \tilde{T}^{z} (\tilde{T}^{zz})^2}{2700} - \frac{14857 \tilde{T}^{z} (\tilde{T}^{zz})^2}{54000 P^2}
    \nonumber\\ 
    && + \frac{26227 \tilde{T}^{z}(\tilde{T}^{zz})^2}{54000 P} + \frac{293 (\tilde{T}^{z})^2
    (\tilde{T}^{zz})^2}{360 P^2} - \frac{7123 (\tilde{T}^{zz})^3}{108000} + \frac{29
    (\tilde{T}^{zz})^3}{8100 P^2} + \frac{29761 (\tilde{T}^{zz})^3}{216000 P}\nonumber\\
    &&  + \frac{2539 P (\tilde{T}^{zz})^3}{259200} - \frac{47 \tilde{T} (\tilde{T}^{zz})^3}{180
    P^2} + \frac{19 \tilde{T}^{z} (\tilde{T}^{zz})^3}{30 P^2} - \frac{131 \tilde{T}^{z}
    (\tilde{T}^{zz})^3}{720 P} + \frac{7 (\tilde{T}^{zz})^4}{360} + \frac{203
    (\tilde{T}^{zz})^4}{1500 P^2} \nonumber\\
&&- \frac{3797 (\tilde{T}^{zz})^4}{36000 P} - \frac{3 \tilde{T}^{z} (\tilde{T}^{zz})^4}{20 P^2}-
\frac{3 (\tilde{T}^{zz})^5}{40 P^2} + \frac{11 (\tilde{T}^{zz})^5}{480  P} +
\frac{(\tilde{T}^{zz})^6}{80 P^2}
\end{eqnarray}
\begin{eqnarray}
\mathcal{F}^{(1,2)}&=&\frac{(\tilde{\deltanew}) (\tilde{\deltanew}^z)}{12}-\frac{17 (\tilde{\deltanew}^z)^2}{120}-\frac{14 (\tilde{\deltanew})^2}{3 P}-\frac{13 (\tilde{\deltanew}) (\tilde{\deltanew}^z)}{6
P}-\frac{113 (\tilde{\deltanew}^z)^2}{120 P}-\frac{211 (\tilde{\deltanew}) \sqrt{z}}{10 \sqrt{P}}\nonumber
\\\nonumber &&-
\frac{71 (\tilde{\deltanew}^z) \sqrt{z}}{20 \sqrt{P}}+\frac{9}{40} (\tilde{\deltanew}) \sqrt{z} \sqrt{P}-\frac{9}{80} (\tilde{\deltanew}^z) \sqrt{z} \sqrt{P}+\frac{(\tilde{\deltanew}^z)^2
P}{24}+\frac{9}{40} (\tilde{\deltanew}^z) \sqrt{z} P^{3/2}+\frac{53 \tilde{T}}{30 P}
\\\nonumber &&
-\frac{17 (\tilde{T}^{z})}{600}+\frac{71 (\tilde{T}^{z})}{300 P}-\frac{25 (\tilde{\deltanew}^z)^2 (\tilde{T}^{z})}{6
P}-\frac{33 (\tilde{T}^{zz})}{5000}+\frac{7 (\tilde{\deltanew}^z)^2 (\tilde{T}^{zz})}{24}-\frac{73 (\tilde{T}^{zz})}{10000 P}+\frac{(\tilde{\deltanew}) (\tilde{\deltanew}^z) (\tilde{T}^{zz})}{2 P}\\\nonumber &&
-\frac{4 (\tilde{\deltanew}^z)^2 (\tilde{T}^{zz})}{3 P}+\frac{7 P (\tilde{T}^{zz})}{5000}-\frac{(\tilde{T}^{z})
(\tilde{T}^{zz})}{20 P}+\frac{3 (\tilde{T}^{zz})^2}{2500}-\frac{3 (\tilde{T}^{zz})^2}{2500 P}+\frac{(\tilde{\deltanew}^z)^2 (\tilde{T}^{zz})^2}{2 P} \\
&& + a_{-1}^{(1,2)} P^{-1} + a_0^{(1,2)} + a_1^{(1,2)} P +a_2^{(1,2)} P^2
\end{eqnarray}

\begin{eqnarray}
\mathcal{F}^{(2,1)}&=&\frac{278 (\tilde{\deltanew})}{375 P^{3/2}}-\frac{(\tilde{\deltanew}^z)}{3000 P^{3/2}}+\frac{1473 (\tilde{\deltanew})}{1000 \sqrt{P}}+\frac{979 (\tilde{\deltanew}^z)}{3600
\sqrt{P}}+\frac{43 (\tilde{\deltanew}) \sqrt{P}}{4500}-\frac{157 (\tilde{\deltanew}^z) \sqrt{P}}{14400}-\frac{1}{600} (\tilde{\deltanew}) P^{3/2}\nonumber \\\nonumber && +
\frac{181 (\tilde{\deltanew}^z) P^{3/2}}{18000}-\frac{1}{600} (\tilde{\deltanew}^z) P^{5/2}+\frac{3 \sqrt{z} \tilde{T}}{4}-\frac{280 (\tilde{\deltanew}) \tilde{T}}{9 P^{3/2}}-\frac{65
(\tilde{\deltanew}^z) \tilde{T}}{9 P^{3/2}}-\frac{211 \sqrt{z} \tilde{T}}{3 P}\\\nonumber &&+
\frac{5 (\tilde{\deltanew}^z) \tilde{T}}{18 \sqrt{P}}+\frac{341 \sqrt{z} (\tilde{T}^{z})}{1200}-\frac{65 (\tilde{\deltanew}) (\tilde{T}^{z})}{9 P^{3/2}}-\frac{3331 (\tilde{\deltanew}^z)
(\tilde{T}^{z})}{900 P^{3/2}}-\frac{287 \sqrt{z} (\tilde{T}^{z})}{20 P}+\frac{5 (\tilde{\deltanew}) (\tilde{T}^{z})}{18 \sqrt{P}}\\\nonumber &&-
\frac{103 (\tilde{\deltanew}^z) (\tilde{T}^{z})}{300 \sqrt{P}}+\frac{29}{240} (\tilde{\deltanew}^z) \sqrt{P} (\tilde{T}^{z})+\frac{261 \sqrt{z} P (\tilde{T}^{z})}{800}-\frac{55
(\tilde{\deltanew}^z) (\tilde{T}^{z})^2}{9 P^{3/2}}+\frac{13 \sqrt{z} (\tilde{T}^{zz})}{2400}\\\nonumber &&-
\frac{253 (\tilde{\deltanew}) (\tilde{T}^{zz})}{450 P^{3/2}}-\frac{1517 (\tilde{\deltanew}^z) (\tilde{T}^{zz})}{1800 P^{3/2}}-\frac{239 \sqrt{z} (\tilde{T}^{zz})}{240 P}+\frac{29
(\tilde{\deltanew}) (\tilde{T}^{zz})}{225 \sqrt{P}}-\frac{419 (\tilde{\deltanew}^z) (\tilde{T}^{zz})}{3600 \sqrt{P}}\\\nonumber &&-
\frac{13}{720} (\tilde{\deltanew}) \sqrt{P} (\tilde{T}^{zz})+\frac{131 (\tilde{\deltanew}^z) \sqrt{P} (\tilde{T}^{zz})}{1200}+\frac{231 \sqrt{z} P (\tilde{T}^{zz})}{1600}-\frac{13}{720}
(\tilde{\deltanew}^z) P^{3/2} (\tilde{T}^{zz})\\\nonumber &&-
\frac{39}{800} \sqrt{z} P^2 (\tilde{T}^{zz})+\frac{5 (\tilde{\deltanew}^z) \tilde{T} (\tilde{T}^{zz})}{3 P^{3/2}}-\frac{9 \sqrt{z} (\tilde{T}^{z}) (\tilde{T}^{zz})}{400}+\frac{5
(\tilde{\deltanew}) (\tilde{T}^{z}) (\tilde{T}^{zz})}{3 P^{3/2}}-\frac{313 (\tilde{\deltanew}^z) (\tilde{T}^{z}) (\tilde{T}^{zz})}{90 P^{3/2}}\\\nonumber &&+
\frac{211 \sqrt{z} (\tilde{T}^{z}) (\tilde{T}^{zz})}{100 P}+\frac{151 (\tilde{\deltanew}^z) (\tilde{T}^{z}) (\tilde{T}^{zz})}{180 \sqrt{P}}+\frac{9 \sqrt{z} (\tilde{T}^{zz})^2}{800}+\frac{29
(\tilde{\deltanew}) (\tilde{T}^{zz})^2}{60 P^{3/2}}-\frac{1537 (\tilde{\deltanew}^z) (\tilde{T}^{zz})^2}{3600 P^{3/2}}\\\nonumber &&+
\frac{299 \sqrt{z} (\tilde{T}^{zz})^2}{800 P}-\frac{(\tilde{\deltanew}) (\tilde{T}^{zz})^2}{15 \sqrt{P}}+\frac{761 (\tilde{\deltanew}^z) (\tilde{T}^{zz})^2}{1800 \sqrt{P}}-\frac{109
(\tilde{\deltanew}^z) \sqrt{P} (\tilde{T}^{zz})^2}{1440}-\frac{9}{400} \sqrt{z} P (\tilde{T}^{zz})^2\\\nonumber &&+
\frac{17 (\tilde{\deltanew}^z) (\tilde{T}^{z}) (\tilde{T}^{zz})^2}{12 P^{3/2}}-\frac{(\tilde{\deltanew}) (\tilde{T}^{zz})^3}{12 P^{3/2}}+\frac{17 (\tilde{\deltanew}^z) (\tilde{T}^{zz})^3}{30 P^{3/2}}-\frac{3
(\tilde{\deltanew}^z) (\tilde{T}^{zz})^3}{20 \sqrt{P}}-\frac{(\tilde{\deltanew}^z) (\tilde{T}^{zz})^4}{8 P^{3/2}} \\
&& + \sqrt{z} \left( a_{-1}^{(2,1)} P^{-1} + a_0^{(2,1)} + a_1^{(2,1)} P +a_2^{(2,1)} P^2 +a_3^{(2,1)} P^3\right)
\end{eqnarray}

\normalsize
\section{Ooguri-Vafa Invariants}
Replacing the generators by their holomorphic limits we can extract the Ooguri-Vafa \cite{Ooguri} invariants from the partition functions. We used for that the conjectured formula in \cite{Walcher1}. It should be noted however that in our formalism the disk invariants $n_d^{(0,1)}$ are extracted from $\frac{1}{2}\mathcal{F}^{(0,1)}$ and the invariants $n_d^{(1,1)}$ are extracted from $2 \mathcal{F}^{(1,1)}$ in order to reproduce the numbers given in \cite{Walcher1}. The clarification of these factors and a better understanding of the multicover formula remains for future work.
\small
\label{tables}
\bigskip
$$
\begin{array}{|r|r|}
\hline
d	&	n_d^{(0,1)} \\ \hline
1	&	30\\ 
3	&	1530\\
5      	&	1088250\\
7       &	975996780\\
9	&	1073087762700\\
11      &	1329027103924410\\
13      &  	1781966623841748930\\
15      &  	2528247216911976589500\\
17      &  	3742056692258356444651980\\
19      &  	5723452081398475208950800270\\
21      &  	8986460098015260183028517362890\\
23      &  	14415044640432226873354788580437780\\
25      &  	23538467987973866346057268850924917500\\\hline
\end{array}
\quad
\begin{array} {|r|r|}
\hline
d	&	n_d^{(0,2)} \\\hline
2	&0\\
4 	&26700\\
6      &38569640\\
8      &58369278300\\
10      &93028407124632\\
12      &153664503936698600\\
14      &260548631710304201400\\
16      &450589019788320352336020\\
18      &791322110332876233623166320\\
20      &1406910190370608901650146628380\\
22      &2526625340233528751485600411725000\\
24      &4575532116961071429530804693412171800\\
26      &8344559227219651245031796423390078968320\\\hline
\end{array}
$$

\medskip

$$
\begin{array}{|r|r|}
\hline
d	&	n_d^{(1,1)} \\\hline
1  &  0 \\
3  &  0 \\
5  &  -2742710 \\
7  & -6048504690 \\
9  & -12856992579490\\
11 & -26585948324529250\\
13 & -54291611312718557630\\
15 & -110080893552894679282680\\
17 & -222191364375273687227005740\\
19 & -447094506460510952531302800200\\
21 & -897635279681074059801246576212490\\
23 & -1799147979326007629352167081015835920\\
25 & -3601314439974327136341483249650915239910\\\hline
\end{array}
\quad
\begin{array} {|r|r|}
\hline
d	&	n_d^{(0,3)} \\\hline
1      &0\\
3      &0\\
5      &117240\\
7      &230877000\\
9      &462884815200\\
11      &915855637274880\\
13      &1804779141114184800\\
15      &3550856539832617041600\\
17      &6982400759593452862593000\\
19      &13728998788327325796353771400\\
21      &26997741895033909653348464555040\\
23      &53102177883967748623102463313529200 \\
25	&104474620947846872117630548142256678000 \\\hline
\end{array}
$$



\begin{thebibliography}{200}
\bibitem{BCOV1}
M.~Bershadsky, S.~Cecotti, H.~Ooguri and C.~Vafa,
{\it Holomorphic anomalies in topological field theories},
Nucl.\ Phys.\  B {\bf 405} (1993) 279,
{\tt [arXiv:hep-th/9302103]}.

\bibitem{BCOV2}
M.~Bershadsky, S.~Cecotti, H.~Ooguri and C.~Vafa,
{\it Kodaira-Spencer theory of gravity and exact results for quantum
string amplitudes},
Commun.\ Math.\ Phys.\  {\bf 165} (1994) 311,
{\tt [arXiv:hep-th/9309140]}.

\bibitem{Witten}
E.~Witten,
{\it Quantum background independence in string theory},
{\tt [arXiv:hep-th/9306122]}.

\bibitem{Yau}
S.~Yamaguchi and S.~T.~Yau,
{\it Topological string partition functions as polynomials},
JHEP {\bf 0407} (2004) 047,
{\tt [arXiv:hep-th/0406078]}.

\bibitem{HKQ}
M.-x.~Huang, A.~Klemm and S.~Quackenbush,
{\it Topological String Theory on Compact Calabi-Yau: Modularity and
and Boundary Conditions},
{\tt [arXiv:hep-th/0612125]}.

\bibitem{Hosono:2007vf}
S.~Hosono and Y.~Konishi,
{\it Higher genus Gromov-Witten invariants of the Grassmannian, and the Pfaffian Calabi-Yau threefolds},
{\tt arXiv:0704.2928 [math.AG]}.

\bibitem{Mayr}
P.~Mayr,
{\it unpublished manuscript},
(2004).

\bibitem{Aganagic}
M.~Aganagic, V.~Bouchard and A.~Klemm,
{\it Topological Strings and (Almost) Modular Forms},
{\tt [arXiv:hep-th/0607100]}.
  
\bibitem{Grimm}
T.~W.~Grimm, A.~Klemm, M.~Marino and M.~Weiss,
{\it Direct integration of the topological string},
{\tt [arXiv:hep-th/0702187]}.

\bibitem{Walcher1}
J.~Walcher,
{\it Extended Holomorphic Anomaly and Loop Amplitudes in Open
Topological String},
{\tt arXiv:0705.4098 [hep-th]}.

\bibitem{Cook}
P.~L.~H.~Cook, H.~Ooguri and J.~Yang,
{\it Comments on the Holomorphic Anomaly in Open Topological String Theory},
{\tt arXiv:0706.0511 [hep-th]}.

\bibitem{Konishi}
Y.~Konishi and S.~Minabe,
{\it On Solutions to Walcher's holomorphic anomaly equations},
{\tt arXiv:0708.2898 [math.AG]}.

\bibitem{Hosono}
S.~Hosono,
{\it Counting BPS states via holomorphic anomaly equations},
{\tt [arXiv:hep-th/0206206]}.

\bibitem{Bonelli}
G.~Bonelli and A.~Tanzini,
{\it The holomorphic anomaly for open string moduli},
{\tt arXiv:0708.2627 [hep-th]}

\bibitem{Walcher2}
J.~Walcher,
{\it Opening mirror symmetry on the quintic},
{\tt [arXiv:hep-th/0605162]}.

\bibitem{Ooguri}
H.~Ooguri and C.~Vafa,
{\it Knot invariants and topological strings},
Nucl.\ Phys.\  B {\bf 577} (2000) 419,
{\tt [arXiv:hep-th/9912123]}.
  
\end{thebibliography}
\end{document}